\documentclass[sigconf,nonacm]{acmart}

\AtBeginDocument{%
  }

\setcopyright{acmlicensed}
\copyrightyear{2018}
\acmYear{2018}
\acmDOI{XXXXXXX.XXXXXXX}

\acmConference[Conference acronym 'XX]{Make sure to enter the correct
  conference title from your rights confirmation emai}{June 03--05,
  2018}{Woodstock, NY}
\acmISBN{978-1-4503-XXXX-X/18/06}

\usepackage{subcaption}
\usepackage{url}
\begin{document}

\title{SenTopX: Benchmark for User Sentiment on Various Topics}


%
\author{Hina Qayyum}
\affiliation{%
  \institution{Macquarie University}
  \city{Sydney}
  \country{Australia}}
\email{hina.qayyum@students.mq.edu.au}

\author{Muhammad Ikram}
\affiliation{%
  \institution{Macquarie University}
  \city{Sydney}
  \country{Australia}}
\email{muhammad.ikam@mq.edu.au}

\author{Benjamin Zhao}
\affiliation{%
  \institution{Macquarie University}
  \city{Sydney}
  \country{Australia}}
\email{ben_zi.zhao@mq.edu.au}

\author{Ian Wood}
\affiliation{%
  \institution{Macquarie University}
  \city{Sydney}
  \country{Australia}}
\email{ian.wood@mq.edu.au}

\author{Mohamad Ali Kaafar}
\affiliation{%
  \institution{Macquarie University}
  \city{Sydney}
  \country{Australia}}
\email{dali.kaafar@mq.edu.au}

\author{Nicolas Kourtellis}
\affiliation{%
  \institution{Telefonica Research}
  \city{Barcelona}
  \country{Spain}}
\email{nicolas.kourtellis@telefonica.com}







\renewcommand{\shortauthors}{Hina et al.}
\begin{abstract}
Toxic sentiment analysis on Twitter ($\mathbb{X}$) often focuses on specific topics and events such as politics and elections. Datasets of toxic users in such research are typically gathered through lexicon-based techniques, providing only a cross-sectional view. his approach has a tight confine for studying toxic user behavior and effective platform moderation. To identify users consistently spreading toxicity, a longitudinal analysis of their tweets is essential. However, such datasets currently do not exist.

This study addresses this gap by collecting a longitudinal dataset from 143K Twitter users, covering the period from 2007 to 2021, amounting to a total of 293 million tweets. Using topic modeling, we extract all topics discussed by each user and categorize users into eight groups based on the predominant topic in their timelines. We then analyze the sentiments of each group using 16 toxic scores. Our research demonstrates that examining users longitudinally reveals a distinct perspective on their comprehensive personality traits and their overall impact on the platform. Our comprehensive dataset is accessible to researchers for additional analysis.
\end{abstract}

\maketitle
\section{Introduction}
An event-centric approach to studying toxic behavior on Twitter involves focusing on specific incidents or events that attract attention and initiate discussions on the platform. User datasets for such analyses are constructed using text or hashtag filters, typically based on individual tweets or a small number of tweets related to the topic. However, the sporadic engagement of users in toxic behavior during these events underscores the importance of addressing the consistency of behavior for fair platform moderation~\cite{app13074550, pavlopoulos2020toxicity}. This necessitates the collection of longitudinal data and the quantification of toxicity over time~\cite{parekh2018studying}, as existing datasets predominantly offer cross-sectional perspectives. To seize this crucial opportunity, this study introduces a longitudinal dataset comprising the timelines of 143K users and 293M tweets spanning the period from 2007 to 2021. Through the application of text modeling techniques, we classify users into eight groups based on prevalent topic categories and examine the distribution of each group's toxicity across all the topics they discuss and the topics they predominately indulge in. We analyze 16 faucets of toxic behavior within these 8 user groups across their tweet timelines. Utilizing longitudinal data will enable the segmentation of user activity, thereby bolstering moderation efforts on the platform and aiding in the identification of users who will exhibit selective or persistent toxicity on Twitter.

Our paper makes the following contributions:
\begin{itemize}
\item We collect a dataset comprising 293 million tweets from 143 users spanning 15 years from 2007 to 2021. We annotate the tweets with topic categories and probability scores for all users. To the best of our knowledge, our dataset is the largest longitudinal Twitter dataset. 
    \item We augment and annotate each tweet in our dataset with a toxicity score using the Perspective API. Each tweet is assigned a probability score of 0-1 for 16 different toxicity categories.
\end{itemize}

To foster future research, we share our dataset with the research community via \url{https://doi.org/10.5281/zenodo.11243662} [Macquarie University IRB Project Reference: 35379, Project ID: 10008, Granted: 27/11/2021]. In the following, we provide details about dataset collection, topic modeling, and toxic sentiment analysis.
\section{Related Work}
Methods for Twitter data collection in toxicity studies often lack proper filtering and validation, compromising reliability. Current approaches are prone to sampling biases~\cite{parekh2018studying}. Social media datasets are often not fully replicable due to platform requirements and the deletion of toxic tweets~\cite{zubiaga2018longitudinal}. The widely used Streaming API only provides real-time data based on specific parameters and does not support historical searches~\cite{cougnon2022collection}. Many Twitter datasets for sentiment analysis lack distinct sentiment annotations for tweets and their topics~\cite{saif2013evaluation}. Recent collection techniques introduce biases, which are investigated through various strategies~\cite{llewellyn2017distinguishing}.

Several topic modeling techniques perform poorly on social network data due to their short and noisy nature, leading to incomparable results across studies~\cite{curiskis2020evaluation}. Egger et al.~\cite{egger2022topic}  
evaluates various document clustering and topic modeling methods using datasets from Twitter and Reddit, highlighting the effectiveness of BERTopic and NMF for Twitter data. Text mining techniques have evolved to extract features from large text corpora, with topic modeling widely adopted~\cite{li2019review, hong2010empirical}. Traditional methods like LDA, LSA, and probabilistic LSA are well-established, while newer algorithms such as NMF, Corex, Top2Vec, and BERTopic are gaining attention~\cite{guo2017mining, albalawi2020using, obadimu2019identifying, sanchez2022travelers}. Despite various algorithms, LDA remains predominant in literature~\cite{gallagher2017anchored}.

Early Twitter studies focused on detecting toxicity~\cite{nobata2016abusive, pavlopoulos2011using, fung2017connected, wulczyn2017ex}, and recent research continues this effort. Taxonomies based on the directness and target of abuse have been proposed~\cite{waseem2017understanding}, along with hierarchical taxonomies~\cite{zampieri2019semeval}. While earlier studies often addressed specific subtypes of toxicity, there is a strong correlation between overall toxicity and its subtypes~\cite{van2018problem}. Models trained to detect general toxicity effectively identify subtypes like hateful language~\cite{pavlopoulos2020toxicity}. However, many datasets lack contextual information for annotators, limiting the effectiveness of toxicity detection methods since conversational context is crucial. 
In this study, we gather a longitudinal dataset and annotate it using 16 models of the Perspective API. We then conduct tweet-level topic modeling using the state-of-the-art BERTweet model to mitigate the biases in Twitter toxicity analyses.

\section{Dataset Collection}
We utilize seven previously published datasets~\cite{gomez2019exploring,kaggle:metoomovement,ribeiro2018like,founta2018large,jha-mamidi-2017-compliment,waseem-hovy-2016-hateful,waseem-2016-racist}, to investigate topics and associated toxic sentiments on Twitter.
Our focus is on English-speaking users, which led us to the acquisition of 143,391 user IDs.
Subsequently, we acquire longitudinal timeline data for each user ID via Twitter's API, we capture the 3,200 most recent tweets (this is due to Twitter API limitations).
After the exclusion of users with fewer than 10 tweets, our dataset comprises 138,430 Twitter users. This dataset spans approximately 293M tweets over 15 years, covering 2007 to 2021. On average, each timeline contains 2,051 total tweets and 1,160 unique tweets, with numerous instances of verbatim repetitions.

\section{Topic modeling}
A \emph{topic model} identifies clusters of words from given documents, which are then interpreted as topics and assigned meaning by humans. After collecting timeline tweets from the users, we employ topic modeling on the text of individual tweets. The resulting topics are then classified to enhance comprehension. Utilizing these classifications, we conduct an extensive toxic sentiment analysis of the user groups indulged in different topic categories. 

This study is particularly significant as it represents the first application of topic modeling to 293 million individual tweets. The following sections detail the specific steps of the topic modeling process applied to our dataset.

\subsection{Contextualized topic model} 
Extracted topics from short, unstructured documents like tweets are often incoherent, making human interpretation challenging. 

Recently, neural topic models have enhanced topic coherence in small documents. These models use contextual embeddings such as BERT~\cite{devlin2019bert} or ELMo~\cite{peters2017semi} to assign weights to words based on their contexts. This technique has advanced the state-of-the-art in neural topic modeling. Consequently, in this work, we employ the contextualized topic model (CTM)~\cite{bianchi-etal-2021-pre} at the tweet level. CTM belongs to a family of neural topic models that take text embeddings as input and produce bag-of-words reconstructions that can be interpreted as models.
For textual embeddings, we leverage BerTweet~\cite{bertweet}, a pre-trained language model for English tweets, trained on 850M English tweets using the RoBERTa pre-training procedure~\cite{https://doi.org/10.48550/arxiv.1904.09482}. According to \cite{conneau-etal-2020-unsupervised}, BerTweet currently outperforms baselines in NLP tasks like classification.
We fine-tune a pre-trained CTM model for our tweets using OCTIS (Optimizing and Comparing Topic models is Simple)~\cite{terragni-etal-2021-octis}, an open-source framework for training, analyzing, and comparing topic models. To evaluate model performance, we use coherence, a standard performance metric. A coherence score ranges from 0 to 1, with higher scores indicating stronger semantic similarity between words in the resulting topics. Our model achieved a coherence score of 0.69. The following subsections explain the steps to perform topic modeling.

\subsubsection{\bf Data preprocessing} 
First, we remove retweets and duplicate tweets from each user's timeline to ensure the topic model processes unique documents more effectively. Next, we standardize the tweets by replacing user mentions and URLs with special tokens. Finally, we exclude users with fewer than 10 tweets, resulting in a dataset of 138,430 users.

\subsubsection{\bf BERTweet textual embedding} 
\label{sec:BERT}
To convert the documents to BerTweet embeddings (vocabulary size 5000),
we first tokenize a 10\% sample of tweets using ``vinai/bertweet-base''~\cite{bertweet}, and preprocess our tweets with CTM preprocessor~\cite{bianchi-etal-2021-pre}. 
Also, during tokenization, we utilize an emoji package to convert emojis into text strings, treating each icon as a single-word token. This process helps capture the context provided by emojis in the tweets.

\subsubsection{\bf Model training} 
To train our model, we utilize a subset comprising 10\% (equivalent to 230 million tweets) of randomly selected unique tweets from Section~\ref{sec:BERT}. Following this, we apply the ``WhiteSpacePreprocessingStopwords'' method within the CTM model to preprocess this random sample of tweets. The specific parameters employed in this step include setting the vocabulary size to 5000, a maximum document frequency of 0.5, a minimum word count of 2, and removing numbers. Finally, we utilize the tokens generated by BERTweet and the preprocessed 10\% of tweets to train our model to extract 200 topics.

\subsubsection{\bf Inferencing}
\label{sec:Inferencing}
We employ the trained model to deduce topics from the remaining tweet documents using the built-in CTM function ``get doc topic distribution''. For each tweet document, the model generates a probability for each extracted topic (in our case, 200 topics). This process is repeated for every tweet, totaling 230,283,810 tweets across 138,430 users, resulting in what we refer to as "Topic probability distribution vectors".

\subsubsection{\bf Hardware} 
The topic models were trained and inference was carried out on an Amazon EC2 {\tt g4dn.8xlarge} instance, equipped with 32 virtual CPUs, 128GB of memory, and a single GPU. The entire process of conducting topic modeling on our dataset took 35 days from beginning to end.

\subsubsection{\bf Interpretation of extracted topics and topic categories} 
\label{sec:pyldavis}
To interpret the topic modeling outcomes, we utilize LDAvis~\cite{pylDAvis}. LDAvis offers metrics for assessing variances among topics~\cite{pylDAvis}, we avail the ``most relevant terms per topic'' principle. LDAvis defines each term's relevance in a topic via a parameter $\lambda$, where $0\leq \lambda \leq 1$, as outlined in Equation \ref{eq:1}:
\begin{equation}
\label{eq:1}
r(w, k \mid \lambda)=\lambda \log \left(\phi_{k w}\right)+(1-\lambda) \log \left(\frac{\phi_{k w}}{p_{w}}\right)
\end{equation}
Here, $\lambda$ governs the emphasis placed on the probability of the term ``w'' within topic ``$k$'' compared to its lift, both measured logarithmically. After experimentation with the relevance parameter, we set $\lambda$ = 0 to exclusively prioritize the top 30 most relevant terms in a topic based solely on their lift. We use these 30 words per topic to label our extracted 200 topics.

\subsubsection{\bf Topic categories}
\label{sec:Topic category}
\begin{table}[t]
    \begin{center}
    \tabcolsep=0.05cm
\scalebox{0.70} {
    \begin{tabular}{l|r|c|l}
    \hline
    \textbf{} & \textbf{\# Users} & \textbf{Discussions} & \textbf{\# Topics} \\
    \textbf{Cat. Label} & \textbf{Discussing Cat.} & \textbf{in Cat.} & \textbf{in Cat.} \\
    \hline
    Everyday & 88,845 (64.1\%) & Pleasantries, wishes, and routine conversations. & 49 (24.5\%) \\
    \hline
    No topic & 29,014 (20.9\%) &  Prepositions, and conjunctions etc. & 43 (21.5\%)\\
    \hline
    News/Blogs & 15,714 (11.3\%) &  News, mentions of newspapers \& News blogs. & 39(19.5\%) \\
    \hline
    Sports & 1,763 (1.3\%) &  Sports events and athletes. & 16(8.0\%) \\
    \hline
    Entertainment & 1,199 (0.9\%) &  Horoscopes, music, TV shows. & 17 (8.5\%) \\
    \hline
    Politics & 1,012 (0.7\%) &  Politicians, political updates, and policies. & 22 (11.0\%) \\
    \hline
    Health/Covid & 597 (0.4\%) & Health-related and Covid-related discussions. & 6 (3.0\%) \\
    \hline
    Cursing & 389 (0.3\%) & Offensive and obscene language or topics. & 8 (4.0\%)\\
    \hline
    \end{tabular}
    }
    \end{center}
    \vspace{-1mm}
    \caption{Overview of topic categories, number of users predominantly discussing each category, main discussions in each category, and the total topics labeled per category.}
    \label{tab:user_groups}
    \vspace{-9mm}
\end{table}
The 200 topics generated by the CTM model lack direct interpretability and exhibit higher-level correlations. To enhance clarity, we manually classify the topics into 8 categories. As discussed earlier, we examine (Section ~\ref{sec:pyldavis}) the 30 most prominent words in each topic, and identify eight primary topic categories: ``Politics'', ``News/Blogs'', ``Health/Covid'', ``Sports'', ``Cursing'', ``Entertainment'', ``Everyday'', and ``No topic''. Next, to label topics with these categories, we assign 3 annotators to label all 200 topics with a category. Any discrepancies in labeling were resolved through discussion and majority voting (requiring agreement from 2 out of 3 annotators). The calculated Fleiss's Kappa of 0.67 indicates substantial agreement among annotators. The specifics of each category, organized by topic counts and proportions, are shown in Table~\ref{tab:user_groups}. This effort provides a clear depiction of all categories and their distribution across the topic

\subsubsection{\bf Normalized global topic probability vector per user}
As explained in Section~\ref{sec:Inferencing}, each tweet yields a probability vector of length 200, with each element representing the probability of participation in one of the 200 topics. Given that users have varying numbers of tweets, we summarize these vectors into an average topic probability vector per user. For instance, if a user has 3,000 tweets, we sum the 200 topic probabilities from each tweet and divide by 3,000 to obtain the user's average topic probability vector. Next, we normalize these user-specific vectors by dividing each by the global topic probability vector. This global vector is computed by averaging the topic probability vectors of all 138K users. This normalization is necessary because certain topics appear with higher probabilities across many tweets. By normalizing, we ensure that prevalent topics do not disproportionately affect the analysis of a tweet or user profile's dominant topics.
This adjustment highlights the importance of using rarely occurring topics and de-emphasizes common topics unless their scores are exceptionally high. These frequent topics typically represent non-essential subjects, as confirmed by analyzing their words and example tweets. Topic models generate such topics to handle common words and vague thematic categories.
\subsubsection{\bf User groups based on engagement across 8 categories}
\label{sec:Thematic diversity}
The establishment of our global topic probability vector that summarizes the topic probabilities for a user and topic categories leads us to divide all users into groups based on the highest engagement in all categories. For this first, we compute \emph{Category probability} for each user and add probabilities of individual topics (from within 200) belonging to each category. For instance ``Everyday'' category has 49 topics so the sum of the normalized global topic probabilities of 49 topics represents the Everyday topic category probability. This vector provides us with a clear way to gauge and compare users' engagement across all categories, and we call it; \emph{Category Probability Vector (CPV)}, represents the probabilities of posting in each of the 8 categories:
\begin{math}
CPV = [CP_{(C_1)},~CP_{(C_2)}~...CP_{(C_8)}].
\end{math}
We divide users into 8 groups based on the highest category probability in this vector. Table \ref{tab:user_groups} details each group.

\vspace{-4.4mm}
\section{Toxicity analysis of user groups}
\label{sec:Topic toxicity}
\begin{figure*}[!ht]
    \begin{subfigure}[b]{.24\linewidth}
    \centering
    \includegraphics[width=\linewidth]{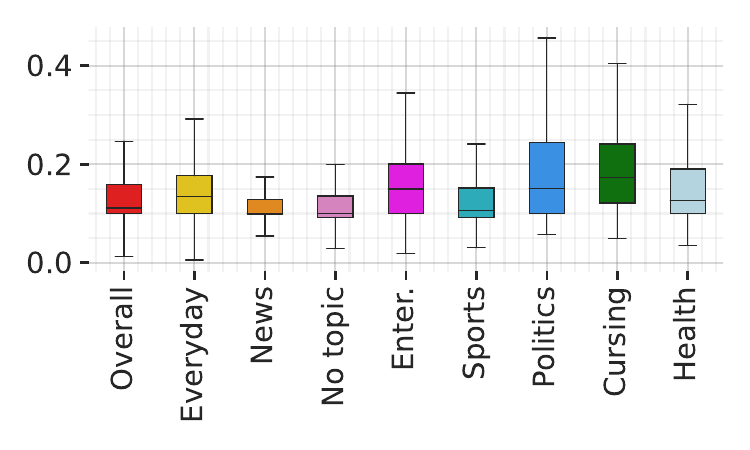}
    \caption{Toxicity}\label{fig:}
    \end{subfigure}
    \begin{subfigure}[b]{.24\linewidth}
    \centering
    \includegraphics[width=\linewidth]{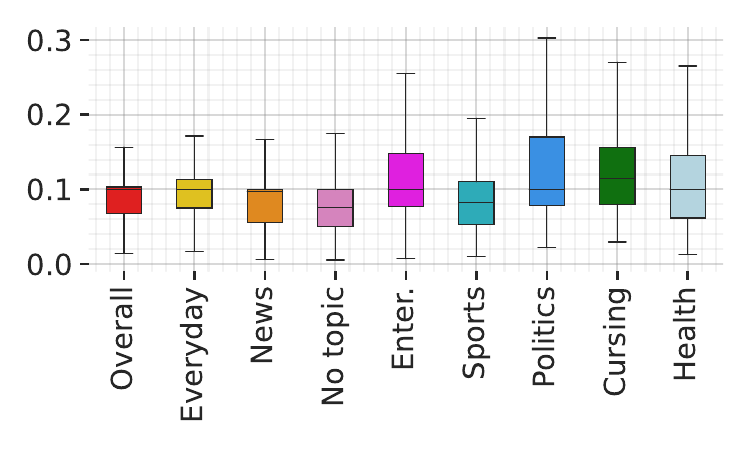}
    \caption{Severe toxicity}\label{fig:}
    \end{subfigure}
     \begin{subfigure}[b]{.24\linewidth}
    \centering
    \includegraphics[width=\linewidth]{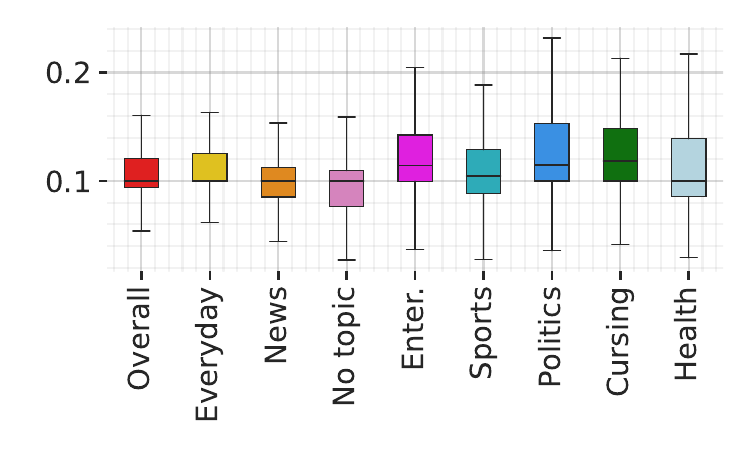}
    \caption{Identity attack}\label{fig:}
    \end{subfigure}
    \begin{subfigure}[b]{.24\linewidth}
    \centering
    \includegraphics[width=\linewidth]{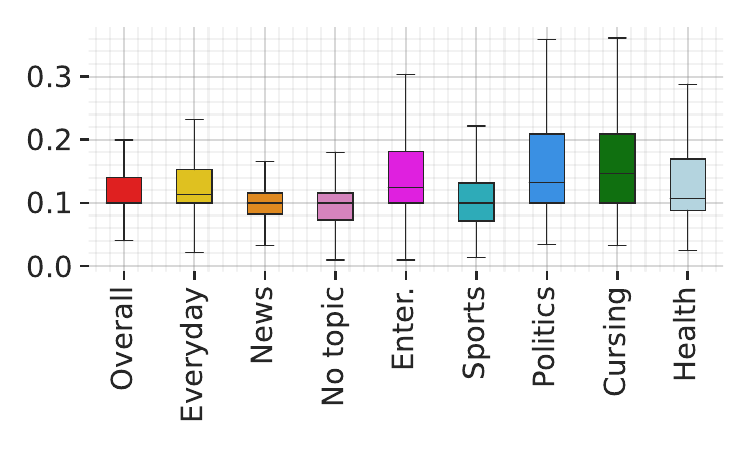}
    \caption{Insult}\label{fig:}
    \end{subfigure}

    \begin{subfigure}[b]{.24\linewidth}
    \centering
    \includegraphics[width=\linewidth]{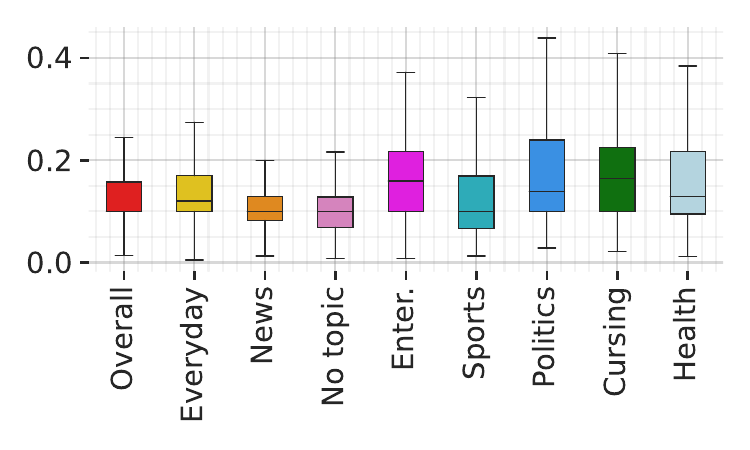}
    \caption{Profanity}\label{fig:}
    \end{subfigure}
    \begin{subfigure}[b]{.24\linewidth}
    \centering
    \includegraphics[width=\linewidth]{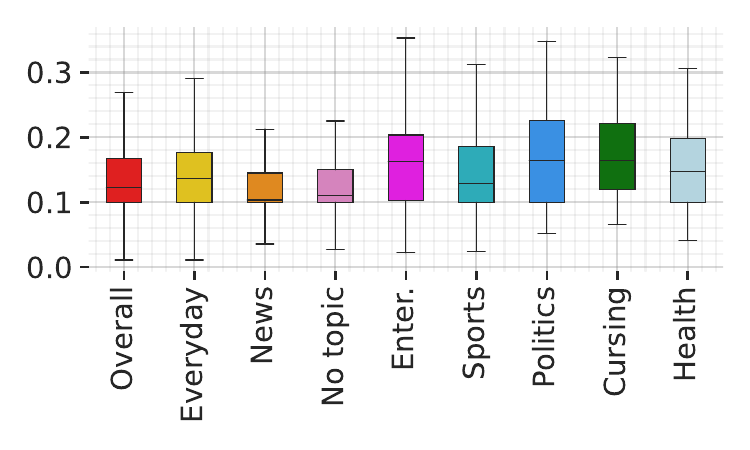}
    \caption{Threat}\label{fig:}
    \end{subfigure}
    \begin{subfigure}[b]{.24\linewidth}
    \centering
    \includegraphics[width=\linewidth]{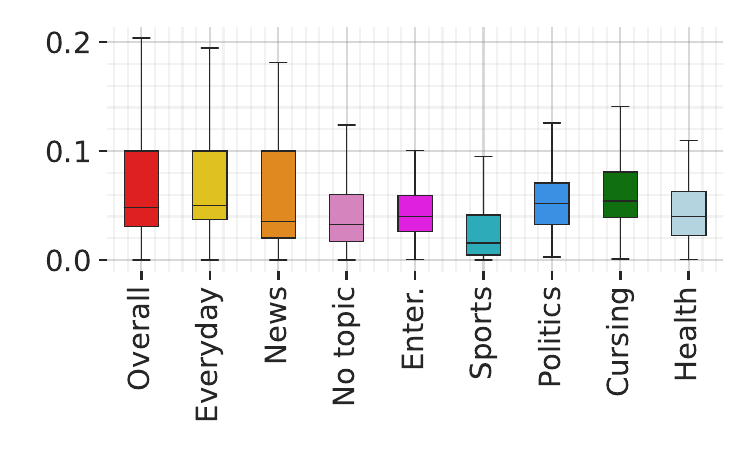}
    \caption{Attack on author}\label{fig:}
    \end{subfigure}
    \begin{subfigure}[b]{.24\linewidth}
    \centering
    \includegraphics[width=\linewidth]{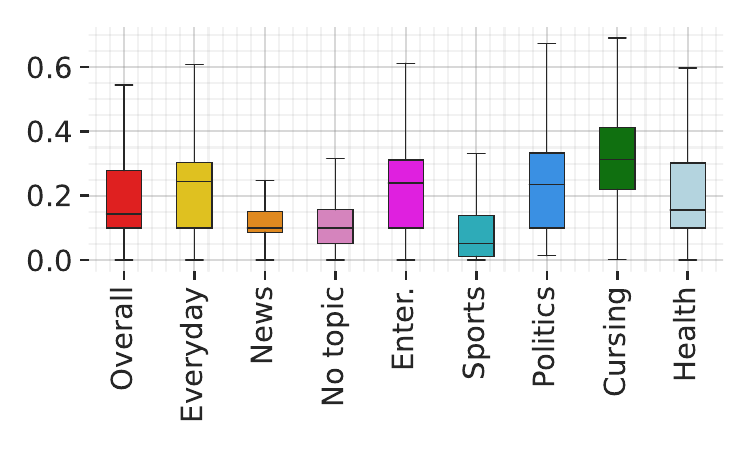}
    \caption{Attack on commenter}\label{fig:}
    \end{subfigure}
    \begin{subfigure}[b]{.24\linewidth}
    \centering
    \includegraphics[width=\linewidth]{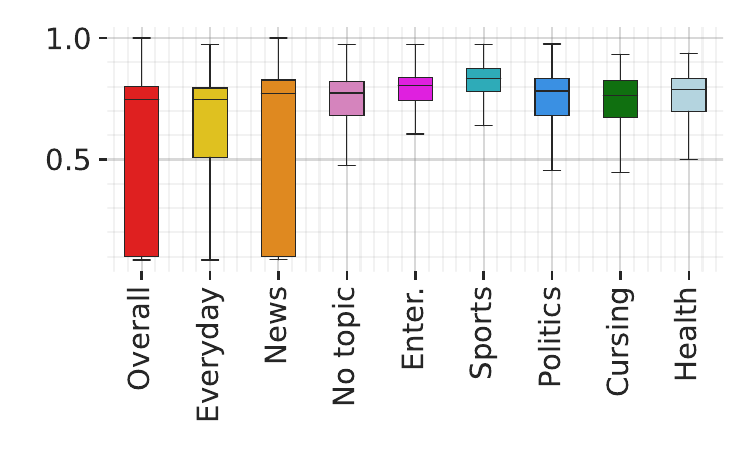}
    \caption{Incoherent}\label{fig:}
    \end{subfigure}
    \begin{subfigure}[b]{.24\linewidth}
    \centering
    \includegraphics[width=\linewidth]{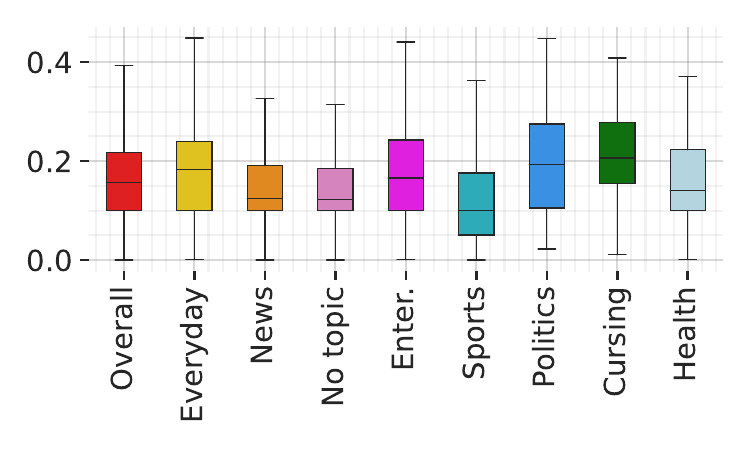}
    \caption{Inflammatory}\label{fig:}
    \end{subfigure}
    \begin{subfigure}[b]{.24\linewidth}
    \centering
    \includegraphics[width=\linewidth]{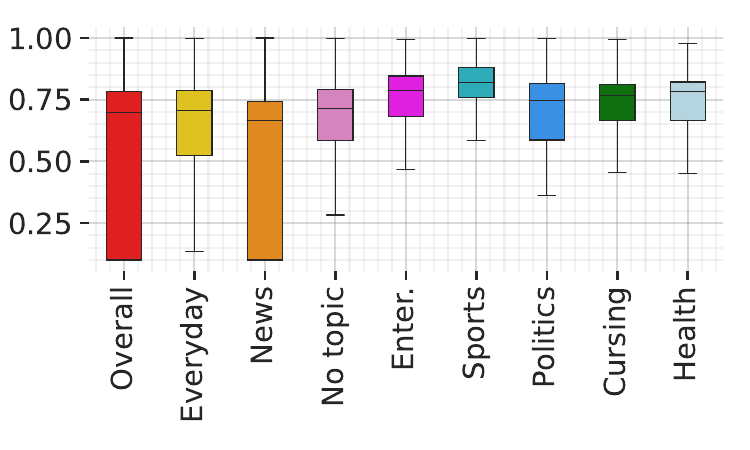}
    \caption{Likely to reject}\label{fig:}
    \end{subfigure}
    \begin{subfigure}[b]{.24\linewidth}
    \centering
    \includegraphics[width=\linewidth]{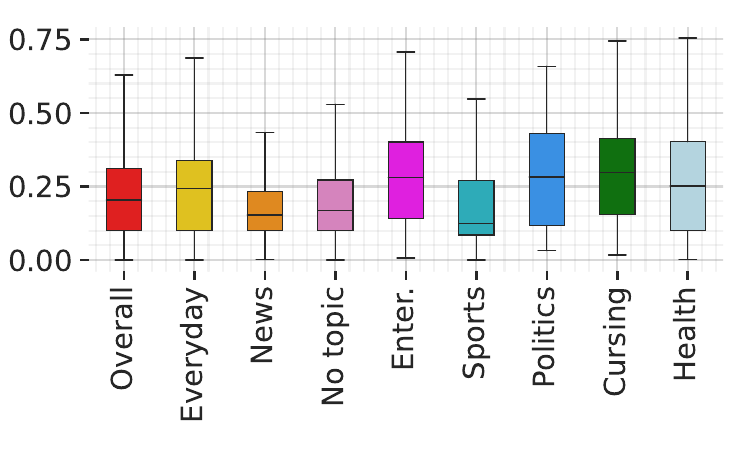}
    \caption{Obscene}\label{fig:}
    \end{subfigure}
    \begin{subfigure}[b]{.24\linewidth}
    \centering
    \includegraphics[width=\linewidth]{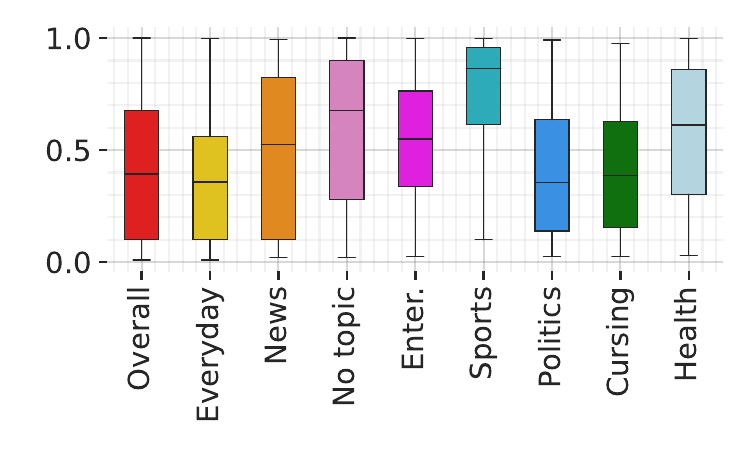}
    \caption{Spam}\label{fig:}
    \end{subfigure}
    \begin{subfigure}[b]{.24\linewidth}
    \centering
    \includegraphics[width=\linewidth]{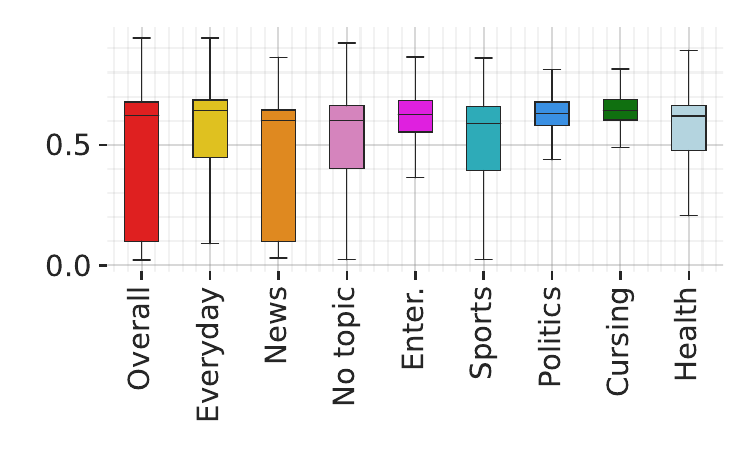}
    \caption{Unsubstantial}\label{fig:}
    \end{subfigure}
    \begin{subfigure}[b]{.24\linewidth}
    \centering
    \includegraphics[width=\linewidth]{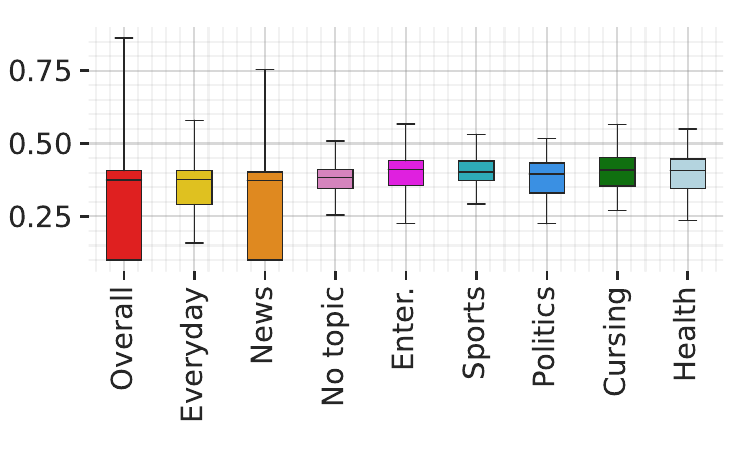}
    \caption{Flirtation}\label{fig:}
    \end{subfigure}
    \begin{subfigure}[b]{.24\linewidth}
    \centering
    \includegraphics[width=\linewidth]{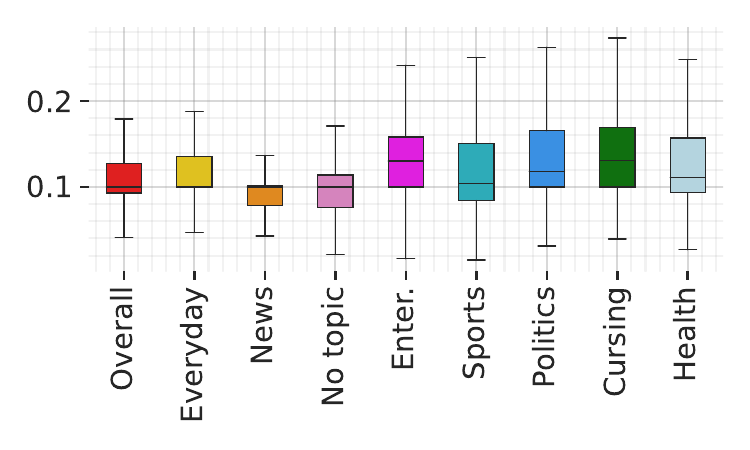}
    \caption{Sexually explicit}\label{fig:}
    \end{subfigure}
   
    \caption{Perspective API models scores; each plot represents mean score per user in each of 8 groups detailed in Section~\ref{sec:Thematic diversity}, while red boxplots in each plot show overall scores.}
    \label{fig:16_plots}
    \vspace{-5mm}
\end{figure*}
We divide our users into 8 groups, each predominantly participating in one of the 8 categories (Table~\ref{tab:user_groups}). 
Predominating longitudinal participation in a particular category reflects the predominant involvement in a category such as politics and sports. 
Next, we conduct a comprehensive analysis focusing on user toxicity on Twitter. We compare all user groups to each other and an overall toxicity score for clarity. To achieve this, we augment each tweet with a toxicity score generated by the \emph{Perspective API}~\cite{perspective}. This API employs 16 distinct machine learning models (details of each model are available via the API~\cite{googleperspectiveapi}) to provide a probability score ranging from 0 to 1, which evaluates the toxicity of the given text. Higher scores correspond to higher levels of toxicity. We utilize the Perspective API models to analyze the toxicity of 293M tweets from all users in our dataset. 

Figure~\ref{fig:16_plots} depicts the Perspective API model scores, with each sub-figure representing the mean score per user in each of the eight groups detailed in (Section~\ref{sec:Thematic diversity}). The x-axis of each plot represents the user groups, including ``Sports'', ``Politics'', ``News/Blogs'', ``Entertainment'', ``Everyday'', ``Health/Covid'', and ``Cursing''. The y-axis represents the probability scores ranging from 0 to 1. Each boxplot provides a visual representation of the distribution of toxicity scores within each group, this allows for a comparison of toxicity levels across different groups in different topic categories. Additionally, the red boxplot in each plot represents the overall result for comparison. This visualization facilitates a clear understanding of the toxicity levels and the distribution of toxicity scores across different topic categories.

We observe that the ``Everyday'' group exhibits a slightly higher toxicity level than the average overall score, with median toxicity scores of 0.15 and 0.1, respectively. This group also demonstrates a propensity for attacking commenters with toxic remarks, as indicated by their high ``likely to reject'' scores. Additionally, their language use often includes slang, abbreviations, and informal conversation, contributing to their incoherent and unsubstantial scores. Finally, this group tends to post fewer comments, as reflected by their low spam scores.

Next, the ``News/Blogs'' group exhibits low levels of toxic behavior. However, their posts are characterized by incoherence and spam, making them unsubstantial and likely to be rejected due to illegibility and spamming. Closer observation reveals that this group primarily shares URLs to news blogs and websites. We also note the similar behavior in ``Sports'' groups.

The "Entertainment" group stands out with high scores across all aspects of toxicity. They exhibit severe toxicity, frequent insults, spamming, and sharing of explicit content, resulting in a higher likelihood of being banned.

Finally, in the ``Politics'' and ``Health'' groups, some outlier users display higher levels of toxic behavior, such as personal attacks and inflammatory posts, without using profanity. Notably, political groups, do not engage in spamming.

It is crucial to highlight that, currently, moderation techniques primarily identify users engaged in spamming as a significant threat. As a result, their posts are deemed unsubstantial and are subject to banning. A holistic approach to studying users through longitudinal quantification of toxicity can shed meaningful light on the overall behavior of users and help us identify users who are involved in other types of toxic behavior.

\section{Conclusion and future directions}
We share our dataset, SenTopX, as a comprehensive benchmark for analyzing user sentiment on Twitter across various topic categories. This mitigates the limitations of previous cross-sectional datasets, we collect a longitudinal dataset comprising 143K Twitter users and 293M tweets from 2007 to 2021. We utilized the contextualized topic model to extract topics from user timelines, categorize topics into 8 categories, and study user groups primarily engaged in each category. Our work provides a novel approach to understanding the dynamics of user engagement and toxic sentiment on Twitter.

We share our detailed dataset to invite further exploration and analysis. Based on the findings of our study, several promising directions for future research can emerge. 
Firstly, an investigation into the long-term effects of user engagement with different toxic categories on social media. This can involve a longitudinal study that tracks users' toxic behavior over extended periods or their entire past posting history.
Secondly, an interesting new direction can involve conducting time series analysis on focus groups to examine the distribution of toxic content around specific events or to predict the groups' future toxic behavior.
Thirdly, finding networks of users inside each category group that we identified and studying the direction of the flow of information in those groups can prove beneficial in identifying the major content generators and distributors. Additionally, studying consistently toxic users' life cycles, including their activity span, bans, and selection of content can be beneficial using our dataset. Furthermore, the SenTopX dataset can be extended to include diverse social media platforms and multi-modal data to enhance the generalizability of the results.

By addressing these areas, future studies can deepen our understanding of user behavior dynamics and inform the development of more effective moderation tools.

\balance
\bibliographystyle{ACM-Reference-Format}
\bibliography{references}

\end{document}